\documentclass[a4paper]{jpconf}
\usepackage{graphicx}
\begin{document}
\title{Deep Learning-Based $^{14}$C Pile-Up Identification in the JUNO Experiment}

\author{Wenxing Fang$^{1*}$, Weidong Li$^1$, Wuming Luo$^1$, Zhaoxiang Wu$^1$, Miao He$^1$}

\address{$^1$ Institute of High Energy Physics, Beijing 100049, People’s Republic of China}

\ead{fangwx@ihep.ac.cn}

\begin{abstract}

Measuring neutrino mass ordering (NMO) poses a fundamental challenge in neutrino physics. To address this, the Jiangmen Underground Neutrino Observatory (JUNO) experiment is scheduled to commence data collection in late 2024, with the ambitious goal of determining the NMO at a 3-sigma confidence level within a span of 6 years. A key factor in achieving this is ensuring a high-quality energy resolution of positrons. However, the presence of residual $^{14}$C isotopes in the liquid scintillator introduces pile-up effects that can impact the positron energy resolution. Mitigating these pile-up effects requires the identification of pile-up events, which presents a significant challenge. The signal from $^{14}$C is considerably smaller compared to the positron signal, making its identification difficult. Additionally, the close event time and vertex between a positron and a $^{14}$C further compound the identification challenge. This contribution focuses on the application of deep learning models for the identification of $^{14}$C pile-up events. It encompasses a range of models, including convolution-based models and advanced transformer models. Through performance evaluation, it shows the deep learning-based methods is promising to 
identify the pile-up events. 

\end{abstract}

\section{Introduction}

\begin{figure}[h]
\begin{center}
\begin{minipage}{15pc}
\includegraphics[width=15pc]{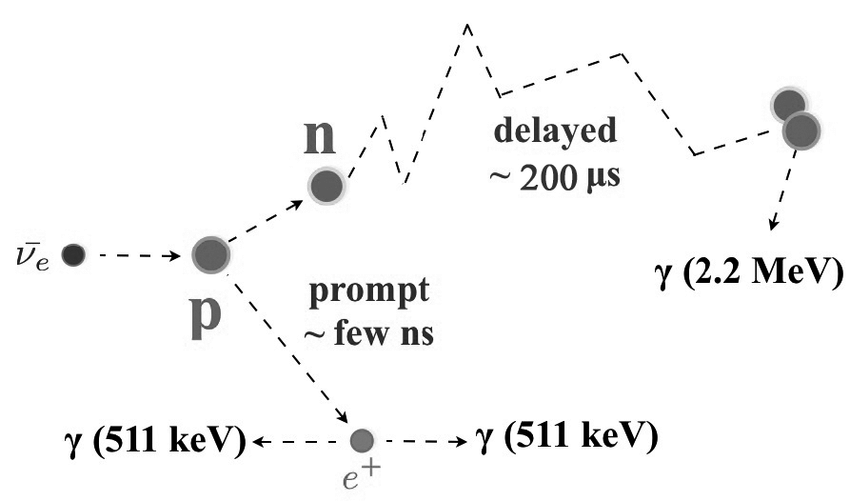}
\caption{\label{fig_IBD}A schematic illustrating the IBD process.}
\end{minipage}\hspace{2pc}%
\begin{minipage}{15pc}
\includegraphics[width=15pc]{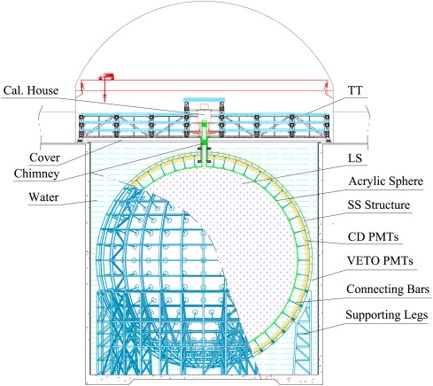}
\caption{\label{fig_JUNO}A schematic view of the JUNO detector.}
\end{minipage} 
\end{center}
\end{figure}

Neutrino physics represents a prominent frontier of current research, with several crucial questions that demand answers. These include inquiries such as the determination of neutrino mass ordering (NMO), the absolute mass of neutrinos, and the intriguing possibility of neutrinos being Majorana particles. By delving into the properties of neutrinos, we can gain a profound understanding of the universe, including phenomena like the asymmetry between matter and antimatter.
The Jiangmen Underground Neutrino Observatory (JUNO) \cite{An_2016_JUNO} is a neutrino experiment designed primarily to measure the NMO. Detecting neutrinos is challenging due to their weak interactions. To enhance the detection cross-section, JUNO employs a 20k ton liquid scintillator (LS) as the sensitive material. The signal process utilized is known as inverse beta decay (IBD), where an antineutrino interacts with a proton, resulting in the production of a positron and a neutron. The positron rapidly annihilates with an electron, generating two 511 keV gamma rays. The neutron, on the other hand, travels a certain distance within the LS before being captured by a hydrogen nucleus, emitting a 2.2 MeV gamma ray. A schematic illustrating the IBD process is presented in Figure \ref{fig_IBD}.

The design of the JUNO detector is illustrated in Figure \ref{fig_JUNO}. The central component is a large acrylic ball containing the 20k ton LS, known as the center detector (CD). Surrounding the CD are numerous photomultiplier tubes (PMTs), including 17,612 20-inch PMTs and 25,600 3-inch PMTs, which are employed to detect the optical photons produced within the LS. Moving outward, there is the water pool Cherenkov detector, serving as an anti-coincidence detector to reject background signals from cosmic-ray muons and providing shielding against surrounding rock-based backgrounds. Positioned at the top is the tracker detector (TT), responsible for detecting cosmic-ray muons.

The careful design of the JUNO detector enables the determination of the NMO with a 3$\sigma$ significance over a 6-year data-taking period. An essential aspect of detector performance is the energy resolution of $e^+$, which needs to be less than 3\%. However, the presence of existing $^{14}$C in the LS can lead to pile-up events with $e^+$, thereby deteriorating the energy resolution. Preliminary results indicate a potential relative degradation of $\sim$2\% at 1 MeV visible energy with a default $^{14}$C activity of 40,000 Bq. This effect is non-negligible and requires mitigation. To address this, the first step involves efficiently identifying pile-up events. Subsequently, a reconstruction algorithm needs to be developed to precisely reconstruct the energy of $e^+$ in pile-up events. This paper focuses on employing deep learning methods for the study of identifying pile-up events.

This paper is organized as follows: Section \ref{sec_Dataset} provides a description of the dataset utilized for model training and testing. Section \ref{sec_2DCNN} outlines the methodology of identifying pile-up events using a two-dimensional Convolutional Neural Network (CNN). Section \ref{sec_1DCNN} presents the approach of identifying pile-up events using a one-dimensional CNN model. Section \ref{sec_Transformer} explains the procedure of identifying pile-up events using a Transformer-based model. The performance results are presented in Section \ref{sec_performance}. Finally, Section \ref{Summary} provides a summary of the findings.


\section{Dataset}
\label{sec_Dataset}

The dataset utilized in this study was generated through simulations using the JUNO offline software (version J22.1.0-rc0). It encompasses Geant4 \cite{Geant4} detector simulation, electronics simulation, waveform reconstruction, and event reconstruction. The dataset consists of two types of events: pure $e^+$ events and $e^+$ events with $^{14}$C pile-up. During training, the kinetic energy of $e^+$ ranges continuously from 0 to 5 MeV. For testing purposes, $e^+$ events with specific kinetic energies of 0, 1, 2, 3, 4, and 5 MeV are employed.

Additionally, it has been observed that for pile-up events with a small number of PMT hits from $^{14}$C (e.g., $\mathrm{nHit_{^{14}C} < 50}$), the impact on the energy resolution is negligible. Therefore, during training, approximately 160,000 pure $e^+$ events and around 90,000 pile-up events (with $\mathrm{nHit_{^{14}C} > 50}$) are utilized.

\section{Deep learning models}

\subsection{Two-dimensional CNN}
\label{sec_2DCNN}

CNN \cite{oshea2015introduction} is a widely utilized deep learning technique known for its ease of understanding and interpretability. In this study, the initial model explored is the two-dimensional CNN (2DCNN). Similar to computer vision problems, the JUNO data is transformed into an image-like format, making it suitable for applying a 2DCNN. To prepare the data for training, the JUNO data is projected onto images with two channels.

\begin{figure}[h]
\begin{center}
\begin{minipage}{10pc}
\includegraphics[width=10pc]{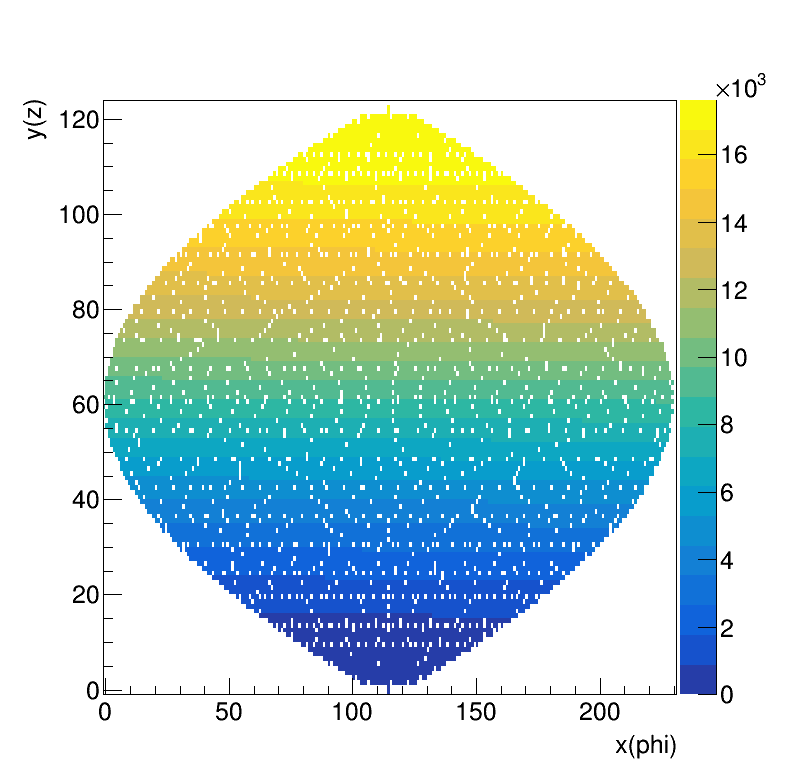}
\caption{\label{fig_ID_map}A mapping between PMT ID (presented by colors) and (x, y) coordinates.}
\end{minipage}\hspace{1pc}%
\begin{minipage}{10pc}
\includegraphics[width=10pc]{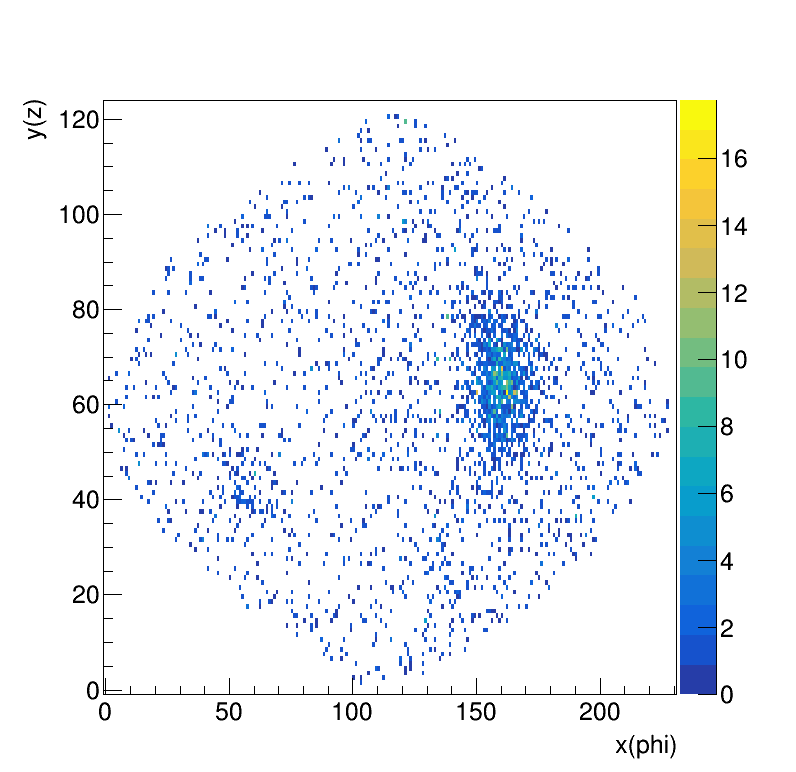}
\caption{\label{fig_chargeByPmts}An example of the charge channel.}
\end{minipage}\hspace{1pc}%
\begin{minipage}{10pc}
\includegraphics[width=10pc]{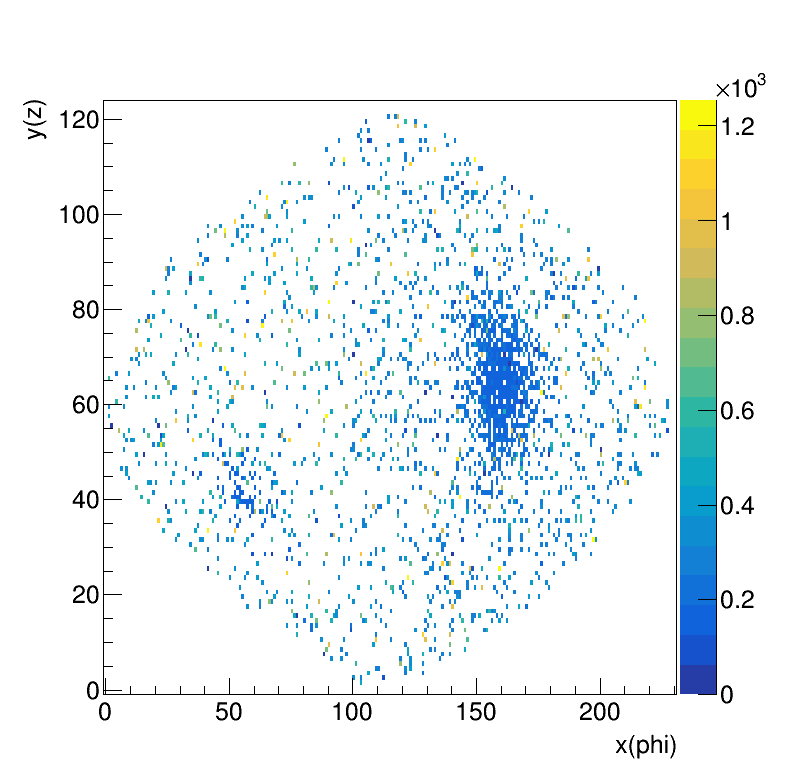}
\caption{\label{fig_ftimeByPmts}An example of the first hit time channel.}
\end{minipage}
\end{center}
\end{figure}

To achieve this, a mapping between PMT ID and (x, y) coordinates (as shown in Figure \ref{fig_ID_map}) is established. Each PMT is then mapped to a corresponding position within the image. For each PMT, there are two measurements available: charge and first hit time. These measurements form the two channels of the image. Examples of the charge channel can be observed in Figure \ref{fig_chargeByPmts}, while the first hit time channel can be seen in Figure \ref{fig_ftimeByPmts}.

To facilitate training, the reconstructed position is rotated to align with the X-axis. Additionally, the values in the charge channel are scaled by a factor of 5, and the values in the first hit time channel are scaled by a factor of 100.

\begin{figure}[h]
\begin{center}
\begin{minipage}{11pc}
\includegraphics[width=10pc]{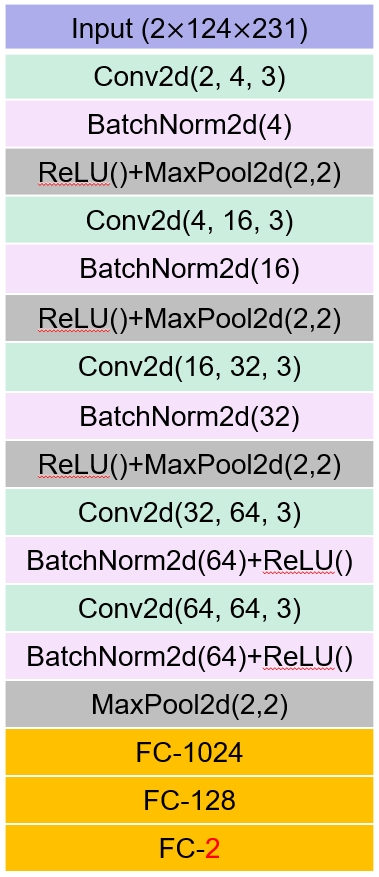}
\caption{\label{fig_2DCNN_model}The architecture of the 2DCNN model.}
\end{minipage}\hspace{1pc}%
\begin{minipage}{11pc}
\includegraphics[width=10pc]{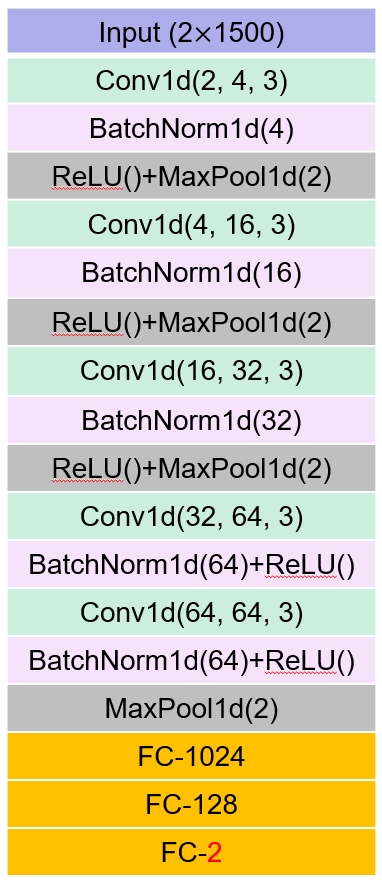}
\caption{\label{fig_1DCNN_model}The architecture of the 1DCNN model.}
\end{minipage}\hspace{1pc}%
\begin{minipage}{12pc}
\includegraphics[width=10pc]{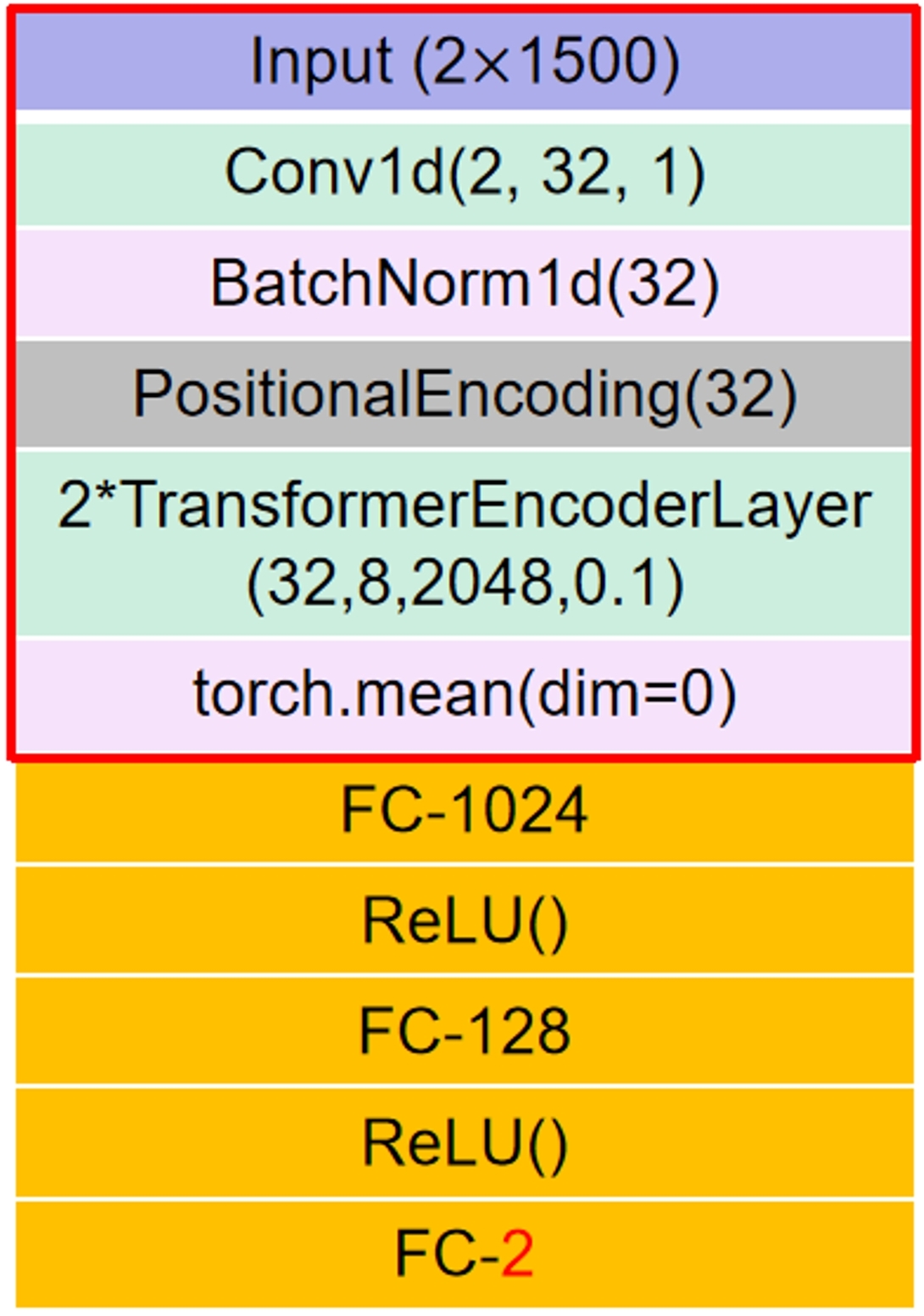}
\caption{\label{fig_TF_model}The architecture of the Transformer-based model.}
\end{minipage}
\end{center}
\end{figure}

The specific architecture of the 2DCNN model is outlined in Figure \ref{fig_2DCNN_model}. It primarily comprises two-dimensional convolutional layers, batch normalization layers, max pooling layers, and fully connected layers. The rectified linear unit (ReLU) activation function is employed throughout the model.

The input data has dimensions of $2\times124\times231$, representing the channel, width, and length of the image, respectively. The model produces two output values, which can be interpreted as the probability of an event being a pile-up event. The cross-entropy loss function is used to train the model, and the Adam optimizer is employed. The learning rate scheduler utilized is the one-cycle learning rate scheduler.

\subsection{One-dimensional CNN}
\label{sec_1DCNN}
For pile-up events, when looking at the number of detector PMT hits versus time, usually there will be two clusters (one from $e^+$, another from $^{14}$C), sometimes just one cluster (when the time between  $^{14}$C and $e^+$ is close). Therefore, one can use the one dimension data (number of PMT hits versus time) to identify the pile-up event.

To identify pile-up events, one can utilize the one-dimensional data representing the number of detector PMT hits versus time. Typically, there will be two clusters in this data: one corresponding to the $e^+$ event and the other to the $^{14}$C event. In some cases, there might be only one cluster when the time between the $^{14}$C and $e^+$ events is close.

In the dataset provided in Section \ref{sec_Dataset}, one-dimensional data with two channels is created for each event.
The first channel represents the number of detector PMT hits versus time, ranging from -250 ns to 1250 ns.
The second channel represents the number of detector PMT hits versus time, with the time of flight subtracted. The time of flight is the flight time of light in the detector from the reconstructed event position to the PMT position. This channel covers the time range from -500 ns to 1000 ns.
To help the training, a simple scaling is applied to both channels. The first channel is scaled by a factor of 10, and the second channel is scaled by a factor of 50.

The architecture of the 1DCNN model is illustrated in Figure \ref{fig_1DCNN_model}. It bears a resemblance to the 2DCNN model, but with a distinction: the 1DCNN model employs one-dimensional neural networks for the convolution, batch normalization, and max pooling layers. The remaining settings, including the loss function, optimizer, and learning rate scheduler, remain consistent with the 2DCNN model. 


\subsection{Transformer-based model}
\label{sec_Transformer}
The Transformer \cite{vaswani2023attention} model has gained significant attention and popularity, especially in the field of natural language processing. This study explores the feasibility of utilizing a Transformer-based model for pile-up identification.

In this approach, the same one-dimensional data used in the 1DCNN model is repurposed as a sequence of vectors. The detailed model structure is depicted in Figure \ref{fig_TF_model}. The data is initially inputted into a convolution layer for embedding. Subsequently, a position encoder is applied, followed by a typical Transformer encoder layer. Finally, several fully connected layers are employed to produce the final output.

The training settings for the Transformer-based model remain the same as those for the 1DCNN model, including the loss function, optimizer, and learning rate scheduler. 


\section{Performance}
\label{sec_performance}

\begin{figure}
\begin{center}
\includegraphics[width=0.32\textwidth]{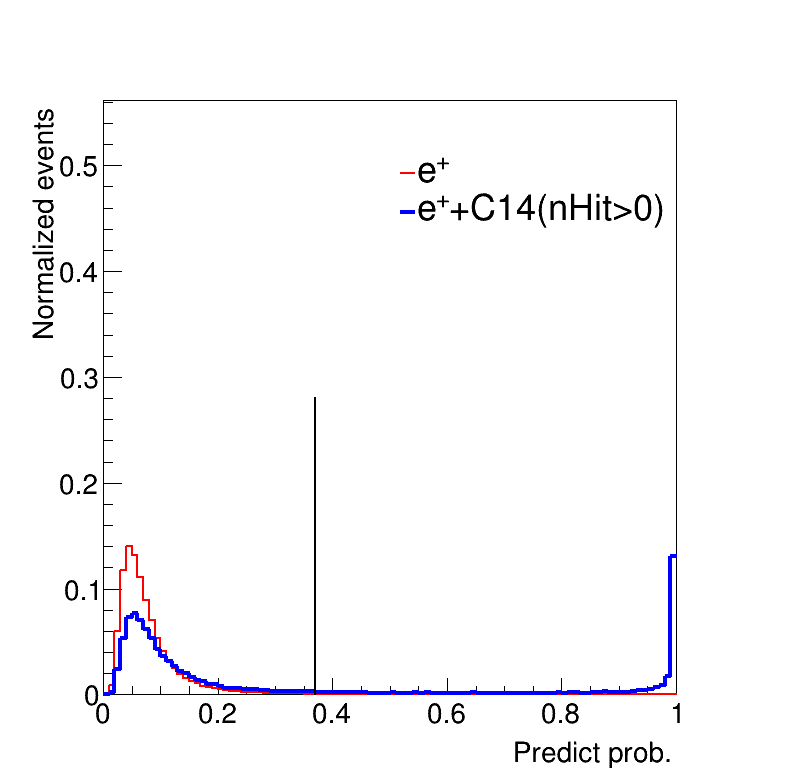}
\includegraphics[width=0.32\textwidth]{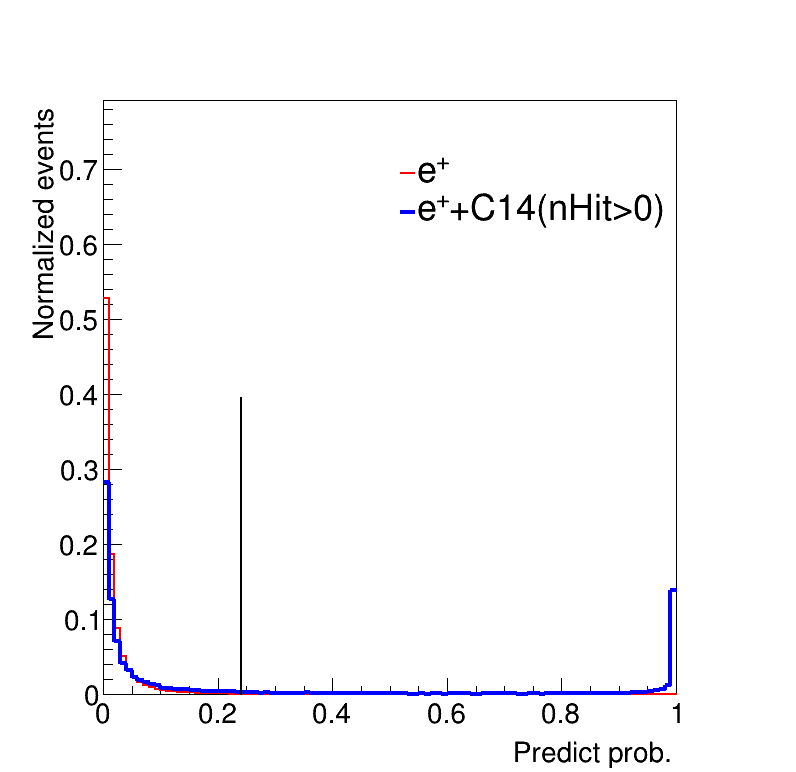}
\includegraphics[width=0.32\textwidth]{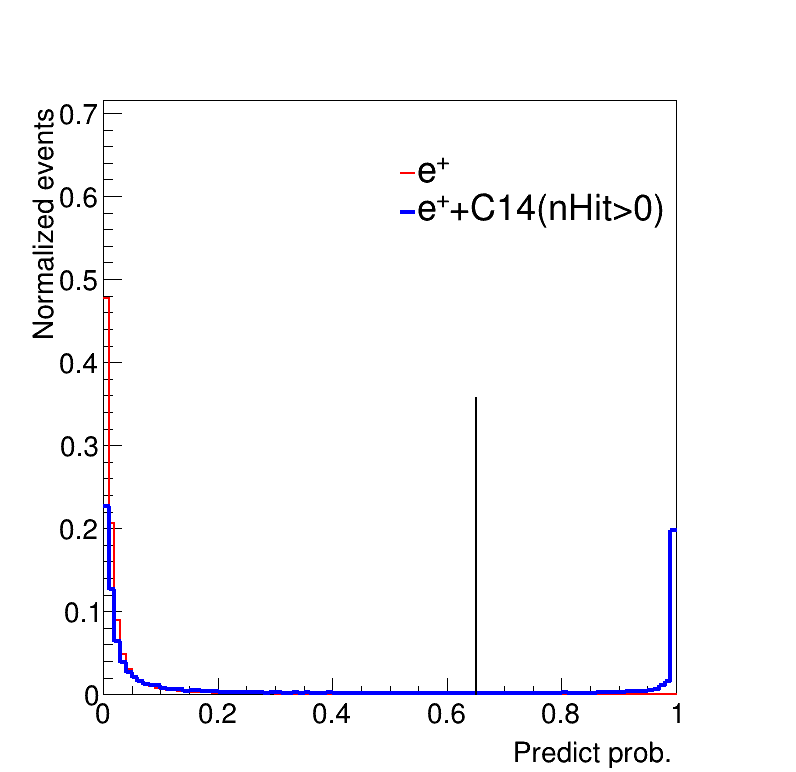}
\caption{The predicted pile-up probabilities for $e^+$ with zero kinetic energy. Red histogram for pure $e^+$ events, blue histogram for pile-up events. The black vertical line represents the 99\% separation point of the red histogram. The left plot is for 2DCNN model, the middle plot is for 1DCNN model, and the right plot is for Transformer-based model.} \label{fig_0MeV_predicted}
\end{center}
\end{figure}

The performance of the three models has been evaluated using testing samples. 
For instance, Figure \ref{fig_0MeV_predicted} illustrates the predicted pile-up probabilities for $e^+$ events with zero kinetic energy. The left plot represents the results from the 2DCNN model, the middle plot corresponds to the 1DCNN model, and the right plot corresponds to the Transformer-based model. Similar results are obtained for other isolated kinetic energy samples (1, 2, 3, 4, and 5 MeV). From Figure \ref{fig_0MeV_predicted}, it can be observed that some pile-up events are not accurately identified. This can be attributed to two main reasons:
1, There are numerous pile-up events with low PMT hits from $^{14}$C (e.g., $\mathrm{nHit_{^{14}C} < 50}$). These types of pile-up events are challenging to identify but have a negligible effect on the energy resolution of $e^+$.
2, Some pile-up events have $^{14}$C positions and times that are close to those of $e^+$. These events are also difficult to distinguish and have a significant impact on the energy resolution of $e^+$.

\begin{figure}
\begin{center}
\includegraphics[width=0.32\textwidth]{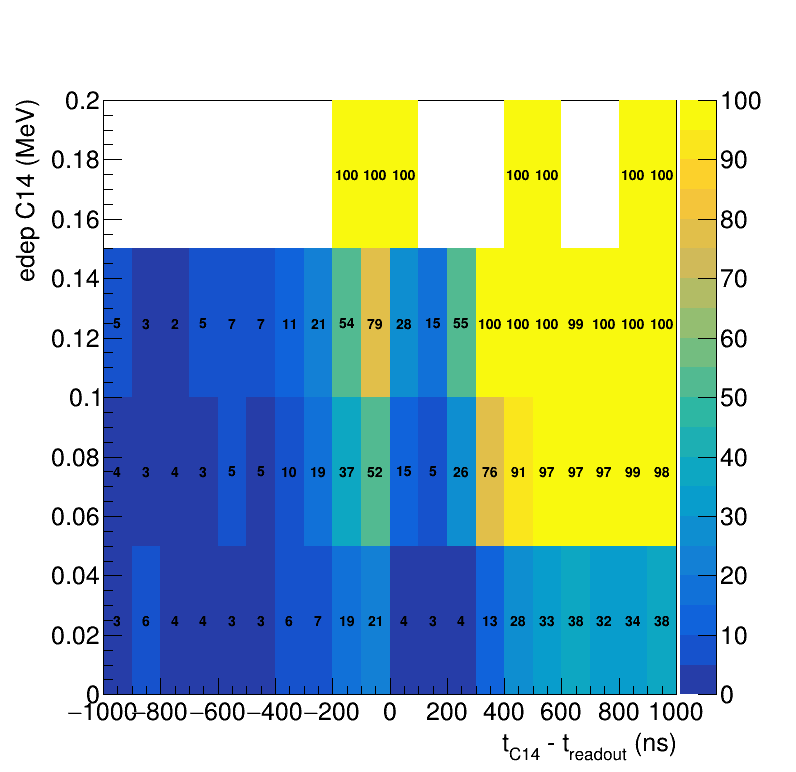}
\includegraphics[width=0.32\textwidth]{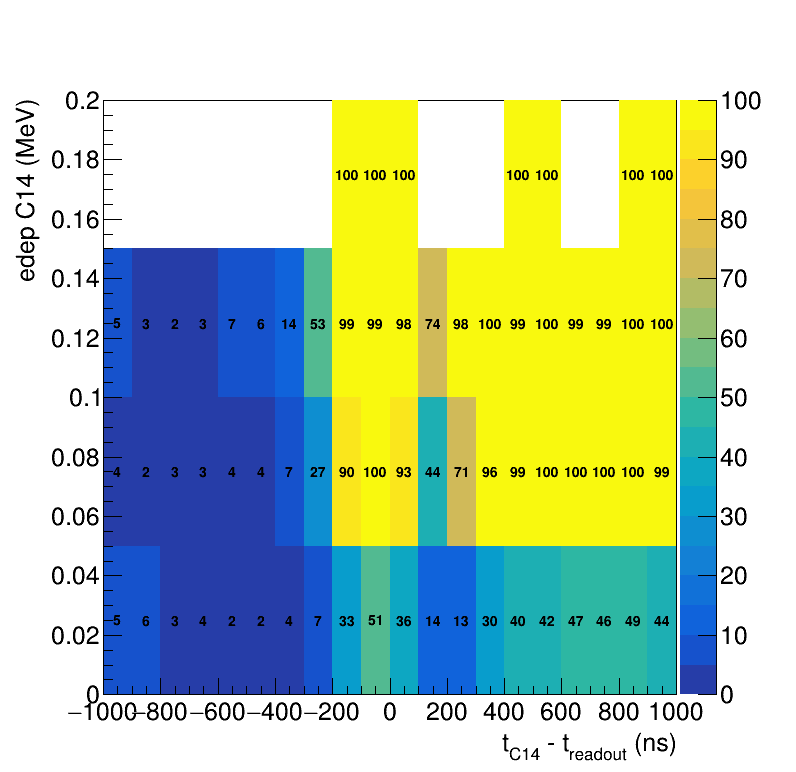}
\includegraphics[width=0.32\textwidth]{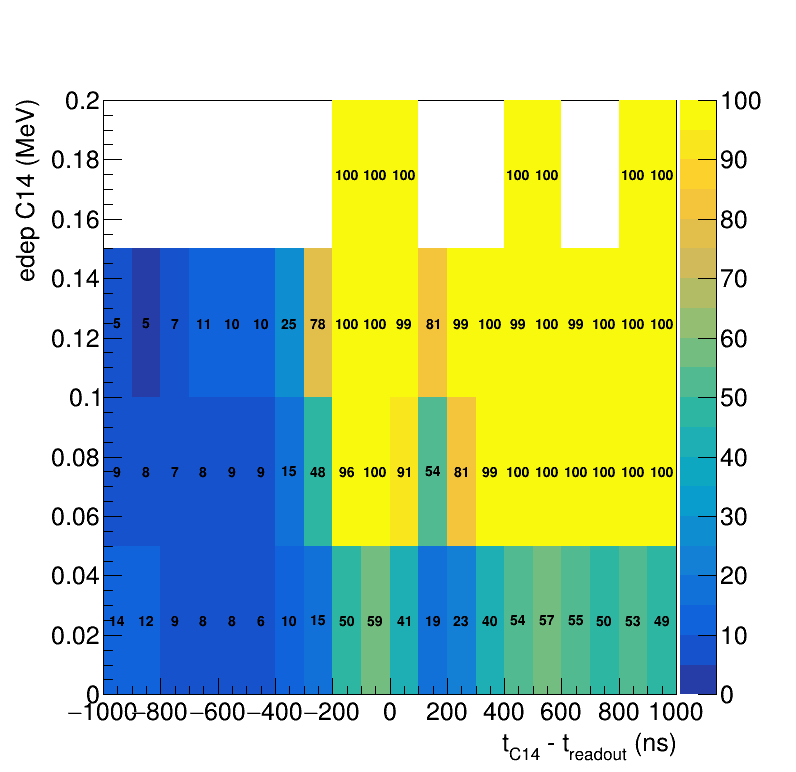}
\caption{The pile-up event identification efficiency for $e^+$ with zero kinetic energy. The efficiency for pure $e^+$ is 99\%. The left plot is for 2DCNN model, the middle plot is for 1DCNN model, and the right plot is for Transformer-based model. X axis is time difference between $^{14}$C and electronics readout. Y axis is the deposited energy of $^{14}$C (for $e^-$ from beta decay of $^{14}$C).} \label{fig_0MeV_eff}
\end{center}
\end{figure}

To assess the performance in detail, Figure \ref{fig_0MeV_eff} illustrates the pile-up event identification efficiency for $e^+$ events with zero kinetic energy. The efficiency for pure $e^+$ events is 99\%. The left plot represents the results from the 2DCNN model, the middle plot corresponds to the 1DCNN model, and the right plot corresponds to the Transformer-based model.

For all models, the identification efficiency for late mixed $^{14}$C events (with a time difference between $^{14}$C and electronics readout larger than 300 ns) is very high. However, in the ``key region" which spanning from 0 ns to 300 ns, the efficiency appears to be suboptimal for the 2DCNN model. In contrast, the 1DCNN model and Transformer-based model exhibit significantly enhanced efficiency in this region.

It is worth noting that other regions with low efficiency have minimal impact on the energy resolution and are of lesser concern. Overall, the performance of the Transformer-based model is similar to that of the 1DCNN model. 

Lastly, the training times for the three models vary. For one epoch, the 1DCNN model takes approximately 4 minutes, the 2DCNN model takes around 60 minutes, and the Transformer-based model takes about 27 minutes.

\section{Summary}
\label{Summary}
The energy resolution of $e^+$ particles holds great importance for the JUNO experiment. The presence of pile-up events caused by $^{14}$C can degrade the energy resolution. Therefore, the identification of pile-up events serves as the initial step in addressing this issue. This paper presents a study of pile-up events identification using different deep learning neural network models.

In comparison to the 2DCNN model, the 1DCNN model demonstrates superior performance in the key region. This improvement is particularly noteworthy as it directly impacts the energy resolution. Furthermore, the Transformer-based model yields similar results to the 1DCNN model, exhibiting promising potential for pile-up event identification.

The successful identification of pile-up events is a crucial advancement towards mitigating the impact of $^{14}$C-induced pile-up on the energy resolution of $e^+$ in the JUNO experiment. By leveraging deep learning techniques and exploring different neural network models, this study provides valuable insights and potential solutions to address this challenge.

\ack
This work is supported by the Innovation Project of the Institute of High Energy Physics under Grant No.E3545BU210, CAS Center for Excellence in Particle Physics.

\section*{References}
\bibliographystyle{iopart-num}
\bibliography{mybib.bib}

\end{document}